\documentclass{iau} 
\usepackage{graphicx}
\usepackage{amssymb}

\title[Modelling energy-dependent pulsar light curves] 
{Modelling energy-dependent pulsar light curves}

\author[Christo Venter \& Monica Barnard]   
{Christo Venter$^1$, Monica Barnard$^1$, Alice K. Harding$^2$ \and Constantinos Kalapotharakos$^{2,3,4}$}

\affiliation{$^1$Centre for Space Research, North-West University, Potchefstroom Campus, \\ 
Private Bag X6001, Potchefstroom 2520, South Africa \\ email: {\tt Christo.Venter@nwu.ac.za } \\[\affilskip]
$^2$Astrophysics Science Division, NASA Goddard Space Flight Center, \\ 
Greenbelt, MD 20771, USA \\
$^3$Universities Space Research Association (USRA), Columbia, MD 21046, USA \\ 
$^4$University of Maryland, College Park (UMDCP/CRESST), College Park, MD 20742, USA}

\pubyear{2017}
\volume{337}  
\jname{Pulsar Astrophysics -­ The Next 50 Years}
\editors{P. Weltevrede, B.B.P. Perera, L. Levin Preston \& S. Sanidas, eds.}

\begin{document}

\maketitle

\begin{abstract}
In recent years, surprise discoveries of pulsed emission from the Crab and Vela pulsars above 100 GeV have drawn renewed attention to this largely unexplored region of the energy range. In this paper, we discuss example light curves due to curvature emission, with good resolution in the different energy bands. Continued light curve modelling may help to discriminate between different emission mechanisms, as well as constrain the location where emission is produced within the pulsar magnetosphere, including regions beyond the light cylinder.
\end{abstract}

\section{Introduction}\label{sec:intro}
Pulsars are broad-band emitters. Their light curves exhibit an intricate structure and evolution with energy, reflecting the various underlying emitting particle populations and radiation components that contribute to this emission as well as the local magnetic field geometry and electric field strength. In addition, Special Relativistic effects modify the emission beam, given the fact that the co-rotation speeds may reach close to the speed of light $c$ in the outer magnetosphere. Much information is encoded in these periodic flashes of intensity.  

One sees additional evolution of the light curves with energy when scrutinising data from ground-based Cherenkov telescopes such as \textit{MAGIC}, \textit{VERITAS}, and \textit{H.E.S.S.-II} that detected pulsed emission from the Crab and Vela pulsars in the very-high-energy (VHE) regime ($>100$~GeV). \textit{MAGIC} recently detected pulsations from the Crab pulsar at energies up to 1~TeV (\cite[Ansoldi et al. 2016]{Ansoldi16}), and \textit{H.E.S.S.-II} detected pulsed emission from the Vela pulsar above 100~GeV, making this only the second pulsar to be detected at these high energies (\cite[de Naurois 2015]{deNaurois15}). Notably, as energy is increased, the main peaks of Crab and Vela seem to remain at the same phase, the intensity ratio of the first to second peak decreases, and the peak widths decrease (\cite[Aleksi\'{c} et al. 2012]{Aleksic12}). Adding data from all energy bands yields an emission spectrum spanning some~20 orders of magnitude (\cite[Harding et al. 2002]{Harding02}, \cite[Abdo et al. 2010a]{Abdo10C}, \cite[Abdo et al. 2010b]{Abdo10V}, \cite[B{\"u}hler \& Blandford 2014]{Buhler14}, \cite[Mignani et al. 2017]{Mignani2017}). 

By constructing detailed physical models, one may hope to disentangle the underlying electrodynamics and acceleration processes occurring in the magnetosphere (see, e.g., the review of \cite[Harding 2016]{Harding16} on using pulsar light curves as probes of magnetospheric structure). As a first approach, we discuss a steady-state emission model that predicts energy-dependent light curves and spectra that result from primary particles emitting curvature radiation (CR). Section~\ref{sec:model} briefly summarises our model, while we present sample light curves in Section~\ref{sec:results}. Conclusions follow in Section~\ref{sec:concl}.

\section{The Model}\label{sec:model}
We use the model of \cite[Harding \& Kalapotharakos (2015)]{Harding15} that assumes a 3D force-free magnetic field (formally assuming an infinite plasma conductivity, so that the electric field is fully screened) as the basic magnetospheric structure. This is a good approximation to the geometry of field lines implied by the dissipative models that assume high conductivity in order to match observed $\gamma$-ray light curves (\cite[Kalapotharakos et al. 2012]{Kalapotharakos12}, \cite[Li et al. 2012]{Li12}, \cite[Kalapotharakos et al. 2014]{Kalapotharakos14}). Both primary particles (leptons) and electron-positron pairs are injected at the stellar surface. The primaries radiate CR and some of these $\gamma$-ray photons are converted into pairs in the intense magnetic fields close to the star. A pair cascade develops, since pairs radiate synchrotron radiation (SR) as well as CR, leading to further generations of particles with lower energies. The primaries are injected with a low initial speed and are further accelerated by a constant electric field (used as a free parameter) in a slot gap (SG) near the last open field lines. In this model, the SG reaches beyond the light cylinder radius $R_{\rm LC}=c/\Omega$, with $\Omega$ the angular speed, up to $r=2R_{\rm LC}$. On the other hand, a pair spectrum is injected at the stellar surface over the full open volume, without any further acceleration taking place. This spectrum is calculated by a steady-state pair cascade code using an offset-polar-cap magnetic field that approximates the effect of sweepback of magnetic field lines near the light cylinder (\cite[Harding \& Muslimov 2011]{Harding11}). The pair multiplicity (number of pairs spawned by each primary particle) is kept as a free parameter to allow for the fact that time-dependent pair cascades may yield much larger values for this quantity (\cite[Timokhin \& Harding 2015]{Timokhin15}) than steady-state simulations (\cite[Daugherty \& Harding 1982]{Daugherty82}). In this paper, we calculate the particle transport as well as CR from primaries. 

\section{Results}\label{sec:results}
\begin{figure}[h!]
  \begin{center}
  \includegraphics[width=0.79\textwidth]{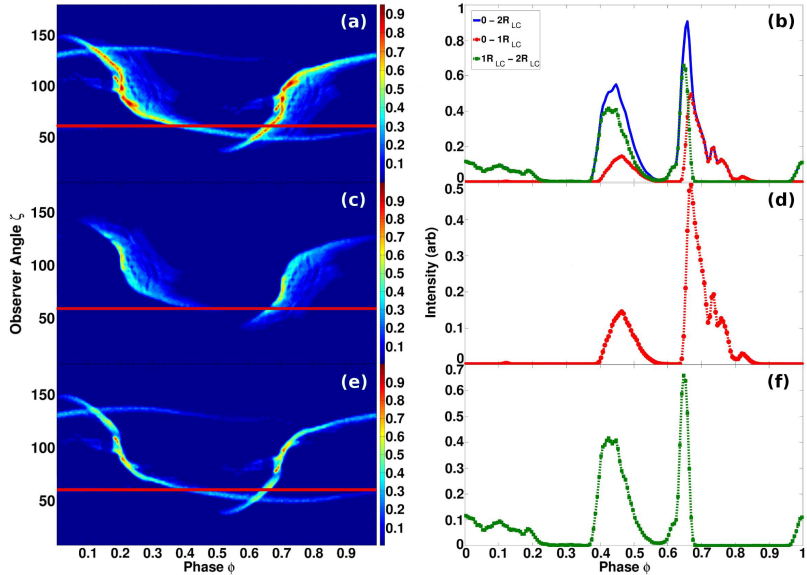}
  \caption{Sky maps (left) and light curves (right) for $100~{\rm MeV}<E_\gamma<10~{\rm GeV}$ and different altitudes $r$, with panels (a) and (b) for $0<r<2R_{\rm LC}$ (solid blue curve), panels (c) and (d) for $0<r<R_{\rm LC}$ (red dashed-circle curve), and panels (e) and (f) for $R_{\rm LC}<r<2R_{\rm LC}$ (green dashed-square curve).}
  \label{fig:Deltar}
  \end{center}
\end{figure}
\begin{figure}[h!]
  \begin{center}
  \includegraphics[width=0.79\textwidth]{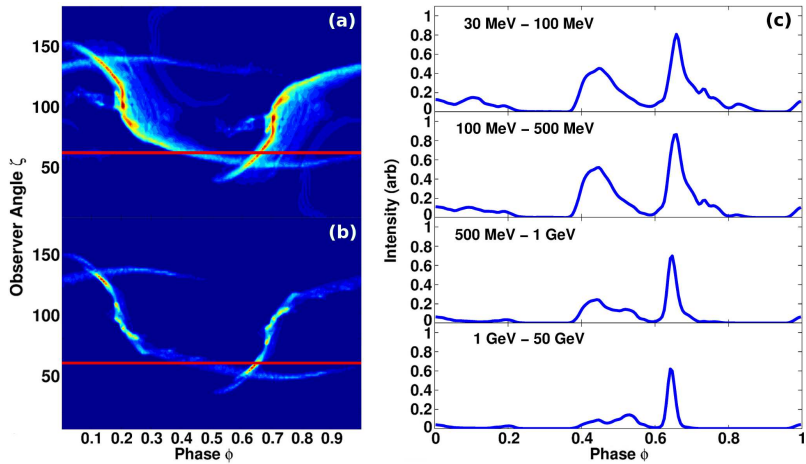}
  \caption{Energy-dependent sky maps (left) and light curves (right) with photon energies $E_{\gamma}$ ranging from (a)~$30~{\rm MeV}<E_{\gamma}<100~{\rm MeV}$, (b)~$1~{\rm GeV}<E_{\gamma}<50~{\rm GeV}$, and (c)~$30~{\rm MeV}<E_\gamma<50~{\rm GeV}$.}
  \label{fig:modelLC}
  \end{center}
\end{figure}

In Figure~\ref{fig:Deltar} we show caustics and light curves for radii $0<r<2R_{\rm LC}$, $r<R_{\rm LC}$, and $r>R_{\rm LC}$, for photon energies $100~{\rm MeV}<E_\gamma<10~{\rm GeV}$. Emission from beyond the light cylinder strongly contributes to the high-energy emission (peaking in the GeV band). One needs quite a fine spatial resolution to obtain smooth light curves. We used 11 rings (self-similar to the polar cap rim) between radial open-volume coordinates $r_{\rm ovc}=0.95$ and $r_{\rm ovc}=0.99$ (\cite[Dyks et al. 2004]{Dyks04}), as well as 180 azimuthal bins. We assumed an inclination angle of $\alpha=45^\circ$ and observer angle of $\zeta=60^\circ$ (both measured with respect to the spin axis).

In Figure~\ref{fig:modelLC} we present sky maps and model light curves for $30~{\rm MeV}<E_{\gamma}<50~{\rm GeV}$,  $\alpha=45^\circ$, and $\zeta=60^\circ$. The light curve morphology changes as $E_{\gamma}$ increases. The first peak's relative intensity decreases with respect to that of the second peak, and the second peak becomes narrower with energy. The second peak position remains roughly constant with energy. This behaviour is qualitatively similar to that observed by \textit{MAGIC} (\cite[Aleksi\'{c} et al. 2012]{Aleksic12}) for the Crab pulsar, and by \textit{Fermi} LAT and \textit{H.E.S.S.-II} for Vela (\cite[Abdo et al. 2010b]{Abdo10V}, \cite[de Naurois 2015]{deNaurois15}). We speculate that this behaviour stems from the fact that the two peaks originate in regions of the magnetosphere that contains magnetic field lines characterised by slightly different radii of curvature $\rho_{\rm c}$. This must be the case since we have assumed a constant electric field in this paper. The spectral cutoff energy should scale with $\rho_{\rm c}^{1/2}$ if the CR reaction limit (when the energy gain and loss rates balance) is reached. Even if this limit is not attained, each peak's spectral cutoff energy should still depend on the local range of $\rho_{\rm c}$ where this emission originates (Barnard et al., in prep.). 

\section{Conclusions}\label{sec:concl}
Modelling of energy-dependent pulsar light curves as well as spectra is vital to disentangle the effects of acceleration, emission, beaming, and magnetic field geometry. We used a 3D emission model assuming CR from primary particles in an SG reaching $2R_{\rm LC}$ to study the evolution of the predicted light curves in different energy bands. We find that emission from beyond $R_{\rm LC}$ (in the current sheet, e.g., \cite[Bai \& Spitkovsky 2010]{Bai10}) constitutes an important contribution to the light curve structure. We also observe that the predicted ratio of the first to second peak intensity decreases, the second peak becomes narrower, and its position in phase remains steady with energy, similar to what has been observed at $\gamma$-ray energies. 

It is not clear what the emission mechanism for high-energy light curves is. The standard models assumed this to be CR (e.g., \cite[Daugherty \& Harding 1996]{Daugherty96}, \cite[Romani 1996]{Romani96}), while newer models focus on SR in the current sheet (\cite[P\'{e}tri 2012]{Petri12}, \cite[Philippov et al. 2015]{Philippov15}, \cite[Cerutti et al. 2016]{Cerutti16}, \cite[Philippov \& Spitkovsky 2017]{Philippov17}). Continued spectral, light curve and now polarisation modelling (\cite[Cerutti et al. 2016]{Cerutti16}, \cite[Harding \& Kalapotharakos 2017]{Harding17}), confronted by quality measurements, may provide the key to discriminate between different models.

\section*{Acknowledgments}
This work is based on the research supported wholly / in part by the National Research Foundation of South Africa (NRF; Grant Numbers 87613, 90822, 92860, 93278, and 99072). The Grantholder acknowledges that opinions, findings and conclusions or recommendations expressed in any publication generated by the NRF supported research is that of the author(s), and that the NRF accepts no liability whatsoever in this regard. A.K.H. acknowledges the support from the NASA Astrophysics Theory Program. C.V. and A.K.H. acknowledge support from the \textit{Fermi} Guest Investigator Program.

\end{document}